\documentclass[aps,prl,twocolumn,showpacs]{revtex4}

\usepackage{graphicx}
\begin{document}
\title{The meaning of S-D dominance}
\author{G.  G. Dussel }
\affiliation{Departamento de F\'{\i}sica Juan Jos\'e Giambiagi, 
Universidad de Buenos Aires and CONICET\\ 1428 Buenos Aires,
Argentina }
\author{H. M. Sofia}
\affiliation{ Departamento de F\'{\i}sica,  Comisi\'on Nacional de Energ\'{\i}a
At\'omica, CONICET and FCEyN, UBA\\
Avda. del Lbertador 8250, 1429 Buenos Aires, Argentina.}

\begin{abstract}
The dominance of S and D pairs in the description of deformed nuclei is one
of the facts that provided sustain to the Interacting Boson Approximation. In
Ref.\cite{DP}, using an exactly solvable model with a repulsive pairing
interaction between bosons it has been shown that the ground state is
described almost completely in terms of S and D bosons. In the present paper
we study the excited states obtained within this exactly solvable
hamiltonian and show that in order to obtain a rotational spectra all the
other degrees of freedom are needed.
\end{abstract}


\maketitle

\section{Introduction}

The interacting boson approximation (IBA)\cite{AI1,AI2,Ot1}
assumes that the elementary excitation used for the description of
the low energy nuclear spectra in open shell nuclei are formed by
pairs of nucleons coupled to angular momentum 0 and 2 (S and D).
These excitations behave in this approximation as bosons. An
intriguing question is how to determine the structure of these S
and D bosons:

It seems natural to use the Hartree-Fock-Bogoliubov (HFB) approximation
where the ground state wave function can be written as in Ref. \cite{ZB}.

\begin{equation}
|\Phi>=N
exp(\frac{1}{2}\sum_{kl}Z_{kl}a_{k}^{\dagger}a_{l}^{\dagger }),
\label{ZB}
\end{equation}
where $a_{k}^{\dagger}$ creates a particle in a state $k$, $N$ is
a normalization constant and $Z$ is the antisymmetric structure
matrix\cite {BM2,Gd1}.

In order to study this point it may be enough to use the results
obtained with schematic calculations using the pairing plus
quadrupole hamiltonian in the Nilsson+BCS approximation. These
calculations show that the S and D bosons are dominant in the
description of the ground state if it is considered a coherent state
of Cooper pairs\cite{Ot2}or as a pair condensate\cite{DDS1}.
However, self consistent HFB calculations for the single
j-shell\cite{BM1} or many shells\cite{BBMV,MVDB} using
unrenormalized simple interactions show that, in these cases, the
space spanned by the S and D pairs does not provide a reasonable
description of well-deformed systems. In Ref. \cite{DDS2} it is
shown that if the interaction strengths are
renormalized to reproduce the value of $\beta$ (deformation parameter) and $
\Delta$ (the superconductive gap) a reasonable description of the
ground state of well-deformed nuclei is obtained.

In Ref.\cite{DES1} a large group of integrable many body
hamiltonians (as well as the constants of motion related to them)
was found. One of these hamiltonians is a repulsive pairing
interaction between bosons with single boson energies that may
have any value. For a particular set of single particle energies
and a repulsive pairing interaction they found that for a large
number of bosons ($N$) the ground state of the system only
contains two types of bosons (in that paper the relevant bosons
were S and P, all other bosons have occupations numbers of order
$\frac{1}{N}$). The reason why it is interesting to consider this
hamiltonian is that the repulsive pairing interaction between the
bosons mucks up the influence of the Pauli principle between the
quasibosons built on fermion pair (see Appendix 1 for a
description of this relation in the single j-shell case). In a
recent letter\cite{DP} a renormalized pairing interaction as well
as a set of single boson´s energies more related to the nuclear
case yields that the S and D states are the only ones that have
large occupations numbers. In Ref.\cite{PDLD} a similar
calculation is done using group theoretical methods (therefore
they are forced to use the pairing interaction instead of the
renormalized one) and the S-D dominance is attributed to the
choice of boson's energies.\\
The present paper uses the recently proposed Richardson-Gaudin
(RG) integrable models \cite{DES1} (for a recent review see
\cite{DPS}) to study the excitations obtained when using a
repulsive pairing interaction between bosons for ``realistic''
situations. By ``realistic'' situations we mean the use of single
boson's energies similar to the ones that correlated pairs of
nucleons have in nuclei. The strength of the pairing interaction
is used as a free parameter. It must be noted that for the single
j-shell the value of the pairing interaction between the
``bosons'' is minus the pairing interaction between fermions, as
shown in Appendix 1.  The use of this treatment allows for the
study of two problems: the first one is the influence of using
realistic values for the boson energies on the S-D dominance as
well as for the interaction strength; the second one is the
possibility of constructing a rotational spectra if a multiboson 
space is considered.

\section{Hamiltonian used and its treatment}

The hamiltonian that will be used is
\begin{equation}
H=\sum_{l} \epsilon_{l} \hat{n}_{l}+\frac{g}{2}
\sum_{kl}(K_{k}^{+}K_{l}^{-}+K_{k}^{-}K_{l}^{+}),
  \label{H1}
\end{equation}
where $\epsilon _{l}$ is the energy of the boson $l$ and the operators 
$K_{i}^{+}=$ and $K_{i}^{-}$ are the generators of an SU(1,1) algebra. The
set of generators $K_{i}^{0}=\frac{1}{2}\hat{n}_{l}+\frac{\Omega _{i}}{4}$, 
$K_{i}^{+}$ and $K_{i}^{-}=(K_{i}^{+})^{\dag }$ satisfies the commutation
relations

\begin{eqnarray}
\left[ K_{i}^{0},K_{j}^{+ }\right] &=&+ \delta _{ij}K_{i}^{+}~ , ~
\left[K_{i}^{0},K_{j}^{- }\right] =- \delta _{ij}K_{i}^{-} ~,  \nonumber \\
& &\left[ K_{i}^{+},K_{j}^{-}\right] =- 2\delta _{ij}K_{i}^{0 }.
\label{KS}
\end{eqnarray}

This SU(1,1) algebra is realized in terms of operators that create
and annihilate bosons. In the usual pseudo-spin representation,
the generators can be written as

\begin{equation}
K_{j}^{0}=\frac{1}{2}\sum_{m}a_{jm}^{\dagger }a_{jm}+\frac{\Omega _{j}}{4}
\text{ \ ,}~~K_{j}^{+}=\frac{1}{2}\sum_{m}a_{jm}^{\dagger }a_{j\bar{m}
}^{\dagger }\text{ \ },
\label{gen}
\end{equation}
where $a_{jm}^{\dagger }\left( a_{jm}\right) $ creates (annihilates) a boson
in the state $|jm\rangle $, $|j\bar{m}\rangle $ is the state obtained by
acting with the time reversal operator on the state $|jm\rangle $, and $
\Omega _{j}$ is the total degeneracy of the single-boson level
$j$. Here the realization involves correlated pair operators. It is
simple to express the pairing hamiltonian in terms of these
operators.

The Hilbert space of $N$ particles moving in shells $k,l,m,....p$ can be
classified in terms of the product of groups SU(1,1)$_{k}\times $SU(1,1)
$_{l}\times $SU(1,1)$_{m}\times ......$SU(1,1)$_{p}$. All the states of the
Hilbert space can be written as
\begin{equation}
|n_{k},...,n_{p,}\nu >=\frac{1}{\sqrt{{\cal N}}}
(K_{k}^{+})^{n_{k}}....(K_{p}^{+})^{n_{p}}|\nu >,
\end{equation}
where \ ${\cal N}$ is a normalization constant. The possible
numbers of pairs in each level is $0\leq n_{s}<N$ and the states
$|\nu >=|\nu _{k},\nu _{l},\nu _{m},...\nu _{p}>$ of unpaired
particles are defined as

\begin{equation}
K_{s}^{-}|\nu >=0,\text{ \ \ \ \ }\hat{n}_{s}|\nu >=\nu _{s}|\nu >,\text{ \
\ \ \ \ }K_{s}^{0}|\nu >=d_{s}|\nu >
\end{equation}
where $d_{s}=\frac{1}{4}(2\nu _{s}+\Omega _{s})$ and $N=2M+\nu ,$
N being the total number of bosons, M the total number of pairs
and \ $\nu $ (usually called the total seniority) the total number
of unpaired bosons that is the sum of all the $\nu_s$.

There are three families of fully integrable and exactly-solvable RG
models that derive from these SU(1,1) algebras, which are referred to as the
rational, trigonometric and hyperbolic models, respectively \cite{DES1}. The
pairing interaction is related to the ``rational'' family, for which they
express the integrals of motion in terms of the above generators as

\begin{eqnarray}
R_{i} =K_{i}^{0}&+&2g\sum_{i ( \neq j)}\left\{\frac{1}
{2(\epsilon_{i}-\epsilon _{j})}
[K_{i}^{+}K_{j}^{-}+K_{i}^{-}K_{j}^{+}]\right.  \nonumber \\
&&~~~~~\left.-~ \frac{1}{ (\epsilon _{i}-\epsilon
_{j})}K_{i}^{0}K_{j}^{0}\right\}. 
\label{cons}
\end{eqnarray}
For each degree of freedom $i$, there is one real arbitrary
parameter that in the rational model is equal to $\epsilon_i$ that
enters the integrals of motion. It can be easily checked that
these operators commute among themselves, and each one commutes
with the operator $K^{0}=\sum_i K^{0}_i$.

The eigenvalue equation $R_i |\Psi_i \rangle = r_i |\Psi_i \rangle$ can be
readily solved using methods analogous to those first introduced by
Richardson to treat the quantum pairing problem \cite{Rich}. The
eigenvectors take the form

\begin{equation}
\left| \Psi \right\rangle =\prod_{\alpha =1}^{M}B_{\alpha
}^{\dagger }\left| \nu \right\rangle ,\quad B_{\alpha }^{\dagger
}=\sum_{l}\frac{1}{(e_{\alpha }-\epsilon _{l})}~K_{l}^{+},
\label{ansa}
\end{equation}
where $|\nu \rangle $ is a state that is annihilated by all the $K_{i}^{-}$ and
$M$ is equal to the number of collective $B^{+}$ operators
that comprise the state. The structures of the collective operators are
determined by a set of $M$ parameters $e_{\alpha }$, which satisfy the set
of coupled nonlinear equations

\begin{equation}
1+4g\sum_{j}\frac{d _{j}}{ 2\epsilon _{j}-e_{\alpha }}+
4g\sum_{\beta \left( \neq \alpha \right) }\frac{1}{e_{\beta
}-e_{\alpha }}=0. \label{Rich}
\end{equation}

The associated eigenvalues take the form

\begin{eqnarray}
r_{i}=d _{i}\left\{ 1-2g\sum_{j\left( \neq i\right) }\frac{d _{j}}
{\epsilon_{i}-\epsilon _{j}} + 4g\sum_{\alpha }\frac{1}{e_{\alpha
}-2\epsilon _{i}}\right\}.
 \label{eigenvalues}
\end{eqnarray}

We note here that each independent solution of the set on nonlinear coupled
equations (\ref{Rich}) defines an eigenstate (\ref{ansa}) and eigenvalues
that are given by

\begin{equation}
E=2g\sum_{\alpha} e_{\alpha}  \label{E}.
\end{equation}

 Following
Richardson \cite{Rich}, the occupation numbers can be calculated
as

\begin{equation}
 \left\langle n_{\lambda}\right\rangle =\left\langle \frac{\partial
H_{P}}{\partial \epsilon_{l}}\right\rangle
=\sum_{\alpha}\frac{\partial e_{\alpha}}{\partial \epsilon _{l}}
~. \label{num}
\end{equation}

From (\ref{H1}), and making use of (\ref{Rich}) and (\ref{E}), a
set of $M$ coupled linear equations in terms of $M$ new unknowns
can be obtained, which give the $\lambda$ boson occupation
numbers. Details of the derivation can be found in ref.
\cite{Rich2}. Dividing these occupation numbers by the total
number of bosons we obtain the occupation probabilities of the
different bosons.

\section{Results and discussion}

In order to relate the calculation to the nuclear case it is
necessary to choose the properties of the boson degrees of freedom
for each angular momentum as well as the coupling constant. The
microscopic image that we have in mind is the Nilsson model and
therefore it seems natural to consider only one type of bosons for
each even angular momentum. For angular momentum two the
seniorities needed to describe the low lying states are well known
\cite{B5D} from the study of the five dimensional harmonic
oscillator. For $\lambda=2$ we will consider only states with
seniority one (J=2), two (J=2 and 4) and three (J=0,3,4,6)). A
similar treatment can be used for the $\lambda$ bosons (the first
excited states corresponding to these degrees of freedom will have
seniority one and angular momentum $J=\lambda$ and are the only
ones that we will consider). It is worthwhile to remark that as
the Hamiltonian has vanishing matrix elements between states with
different seniorities, the states connected by the hamiltonian
must have the same seniorities. There are too many parameters if
the boson's single particle energies are left as free parameters.
We performed all our calculation with a large but meaningful
number of bosons (40) and we choose the single boson energies  by
imposing the condition that the ratios of calculated energies of
the first $2^+$, $4^+$, $6^+$ and $8^+$ states are as close as
possible to the values provided by the rigid rotor for
$G/h\nu_2=.72$ (for this value of $G/h\nu_2$ the ratios of the
energies of the ground state band have reached what we can
consider their asymptotic values) taking into account bosons with
all the even angular momentum up to 12. As we do not know the
ground state energy, we can only determine the differences between
the bosons single particle energies. We define our energy scale by
making the difference of energy between the $\lambda=2$ and
$\lambda=0$ bosons equal to one. The resulting energy differences
(to the J=0 boson) have values which are reasonables: E$_4=1.25$;
E$_6=1.50$; E$_8=2.12$; E$_{10}=2.28$ and ; E$_{12}=2.42$.

 In Fig. 1 we show the excitation energy of the $2_1$ state
as function of $G/h\nu_2$ (being $h\nu_2$ the energy of the
$\lambda=2$ boson).
 The asymptotic value ($\approx .124$) provides us with an
 absolute energy scale. If this state is the $2_1$
 state of the ground state rotational band the energies are of the
 right order of magnitude and $G/h\nu_2$ must have a value of the order
 of 0.2 to reproduce the fermionic strength.

\begin{figure}

\includegraphics[ height=6.5cm, width=5.2cm]{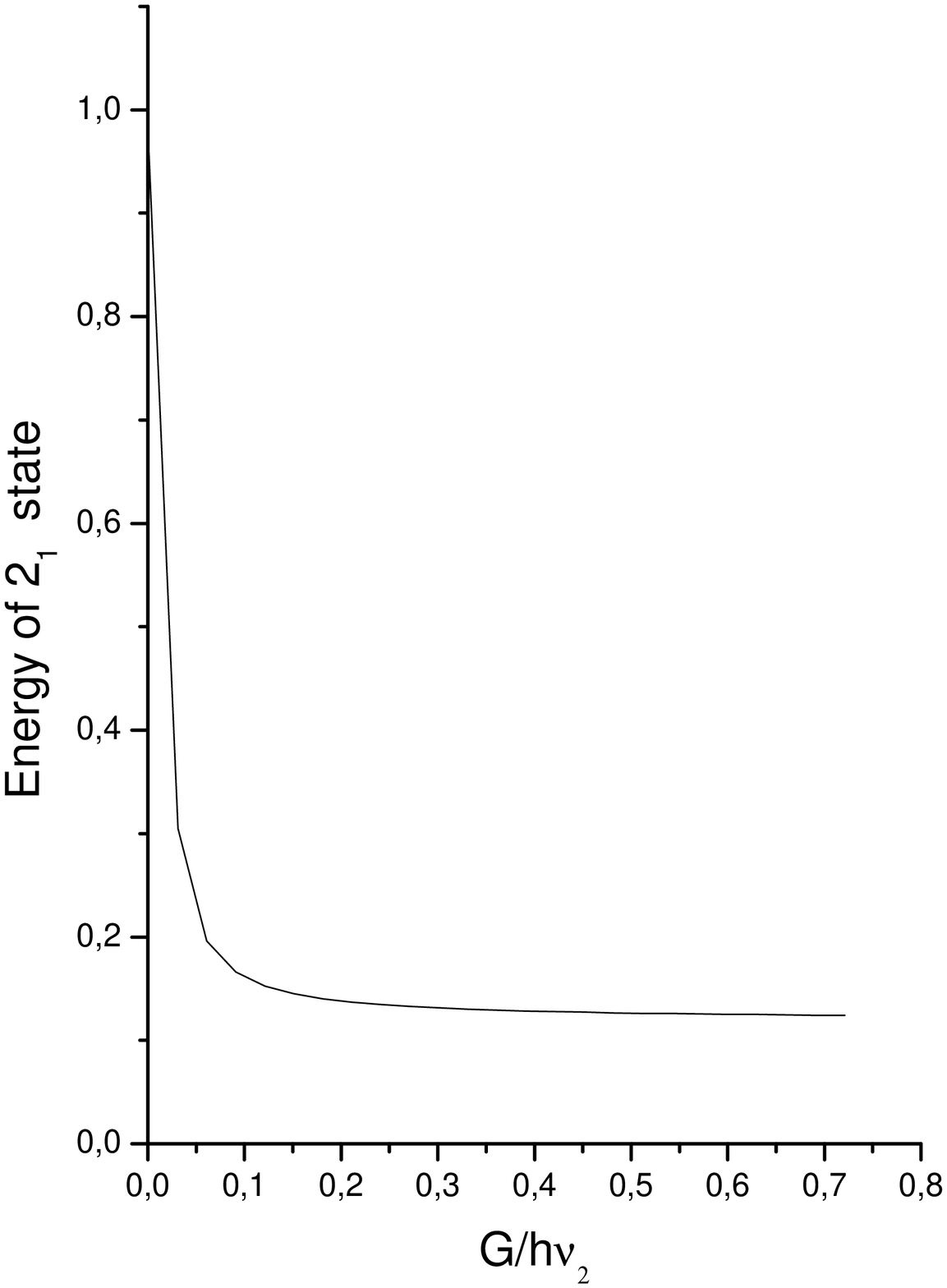}

\caption{Energy of the $2_1$ state as function of
the scaled strength parameter $G/h\nu_2$.}
\label{fig1}
\end{figure}

Initially we include all bosons with even angular momentums up to
$\lambda=12$.
 In Fig. 2 we show the ratio between the
 energy  of the first state for each even angular momentum (up to J=12) to the
 $2_1$ state as function of $G/h\nu_2$. It is evident the
 departure from the rigid rotor values when the angular momentum
 increases or $G$ decreases. In the region of $G/h\nu_2 \approx 0.2$ the structure
  of the excitations is similar
 to the rigid rotor. For the model we are using, the states are
 labeled by the seniority of the different bosons. To get a feeling of the types
 of result obtained in table 1 we list the excitation energies for 
$G/h\nu_2 =0.72$ for some  selected sates, showing also the possible 
angular momentum of these states.
  It follows that there are many excited levels in the low energy region and that
  it is difficult to recognize the excited
  rotational bands.

\begin{table*}[h]\centering
\caption{Excitation energies, in units of $h\nu_2$, for the three
first solutions with the seniorities displayed in the table. A
list of all the angular momentum related with this seniority
structure are given in the last column.}
\begin{tabular}{cccccccccc}
$\tilde{M}$ &$\nu_{J=0}$ &$\nu_{J=2}$ &$\nu_{J=4}$ &$\nu_{J=6}$
&$\nu_{J=8}$ &$E_1$ &$E_2$ &$E_3$
&$Possible~angular~momentums$\\
 \tableline
 20 & 0 & 0 & 0 & 0 & 0 & 0.00000 &0.59667 &1.35124 &$0$ \\
 19& 1& 1& 0 & 0& 0& 0.12424& 0.79629&1.51183&$2$ \\
 19& 1& 0& 1 & 0& 0& 0.41361& 0.99778&1.79228&$4$ \\
 19& 1& 0& 0 & 1& 0& 0.86824& 1.46588&2.18733&$6$ \\
 19& 1& 0& 0 & 0& 1& 1.54591& 2.14410&2.88702&$8$ \\
 19& 0& 2& 0 & 0& 0& 0.29651& 1.03966&1.72393&$2,4$ \\
 18& 1& 3& 0 & 0& 0& 0.51105& 1.32094&1.98002&$0,3,4,6$ \\
 18& 1& 2& 1 & 0& 0& 0.75453& 1.50430&2.21865&$0,1,2^2,3^2,4^2,5^2,6^2,7,8$ \\
 18& 1& 2& 0 & 1& 0& 1.19265& 1.95006&2.59798&$2,3,4^2,5^2,6^2,7^2,8^2,9,10$ \\
 18& 0& 3& 1 & 0& 0& 0.993228& 1.81334& 2.50029&$0,1^2,2^3,3^3,4^4,5^3,6^3,7^3,8^2,9,10$\\
\end{tabular}
\label{table1}
\end{table*}

\begin{figure}
\includegraphics[ height=6.5cm, width=5.2cm]{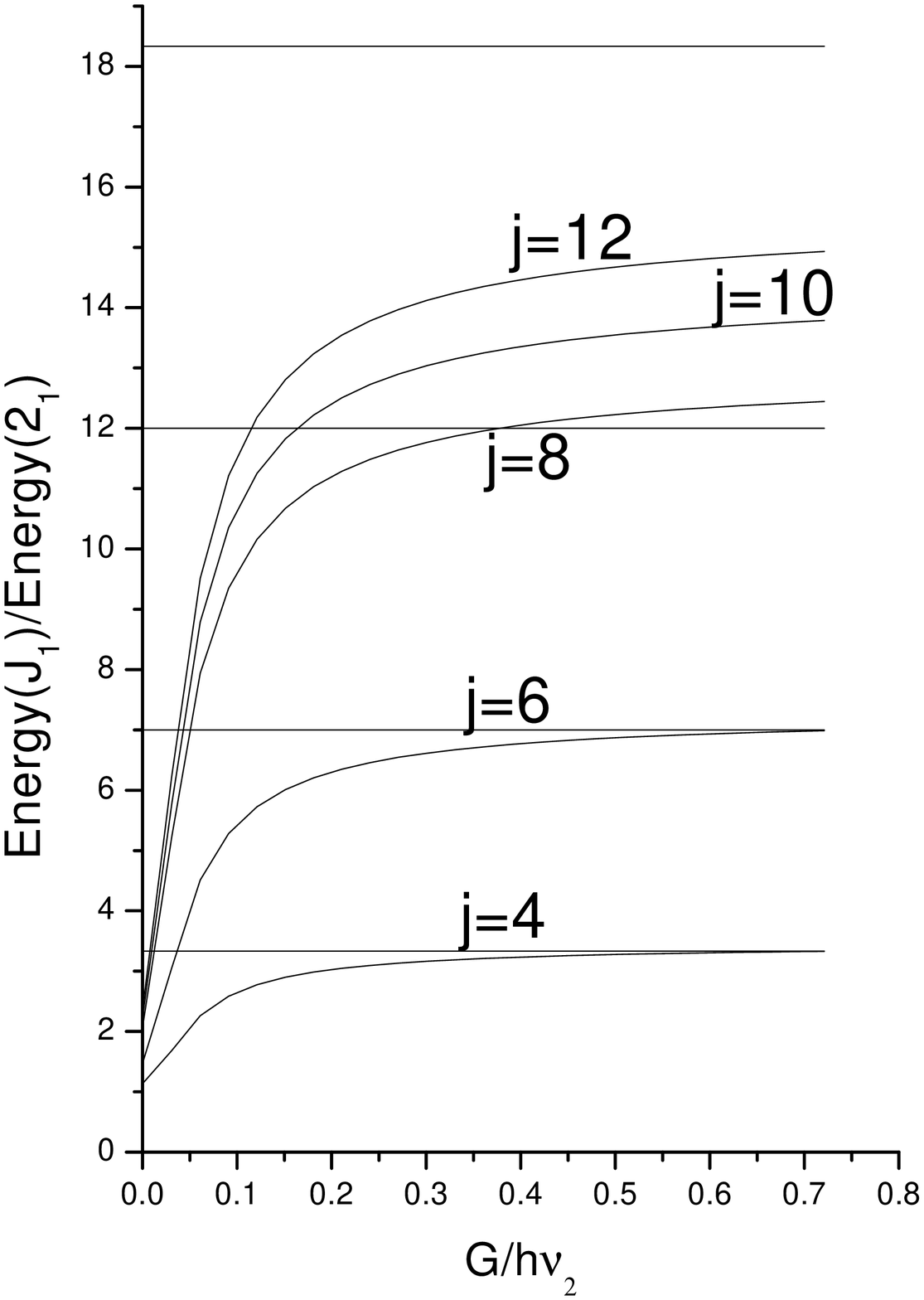}
\caption{Ratio of the energy of the $J_1$ states to
the energy of the $2_1$ state as function of the scaled strength
parameter $G/h\nu2$ for a calculation that includes all even
values of J up to 12. The rigid rotor values are shown to guide
the eye.}
\label{fig2}
\end{figure}

  In Fig. 3 we show the occupation probabilities of the different boson levels
obtained  for some selected states. In general only two boson
levels are populated (the S and D bosons share more than 85\% of
the occupation probabilities). The contribution of the $\lambda=0$
boson will be more important if we increase the number of bosons
or if we consider a renormalized interaction as the one used in
Ref.\cite{DP}.

\begin{figure}
\includegraphics[ height=10.5cm, width=7.5cm]{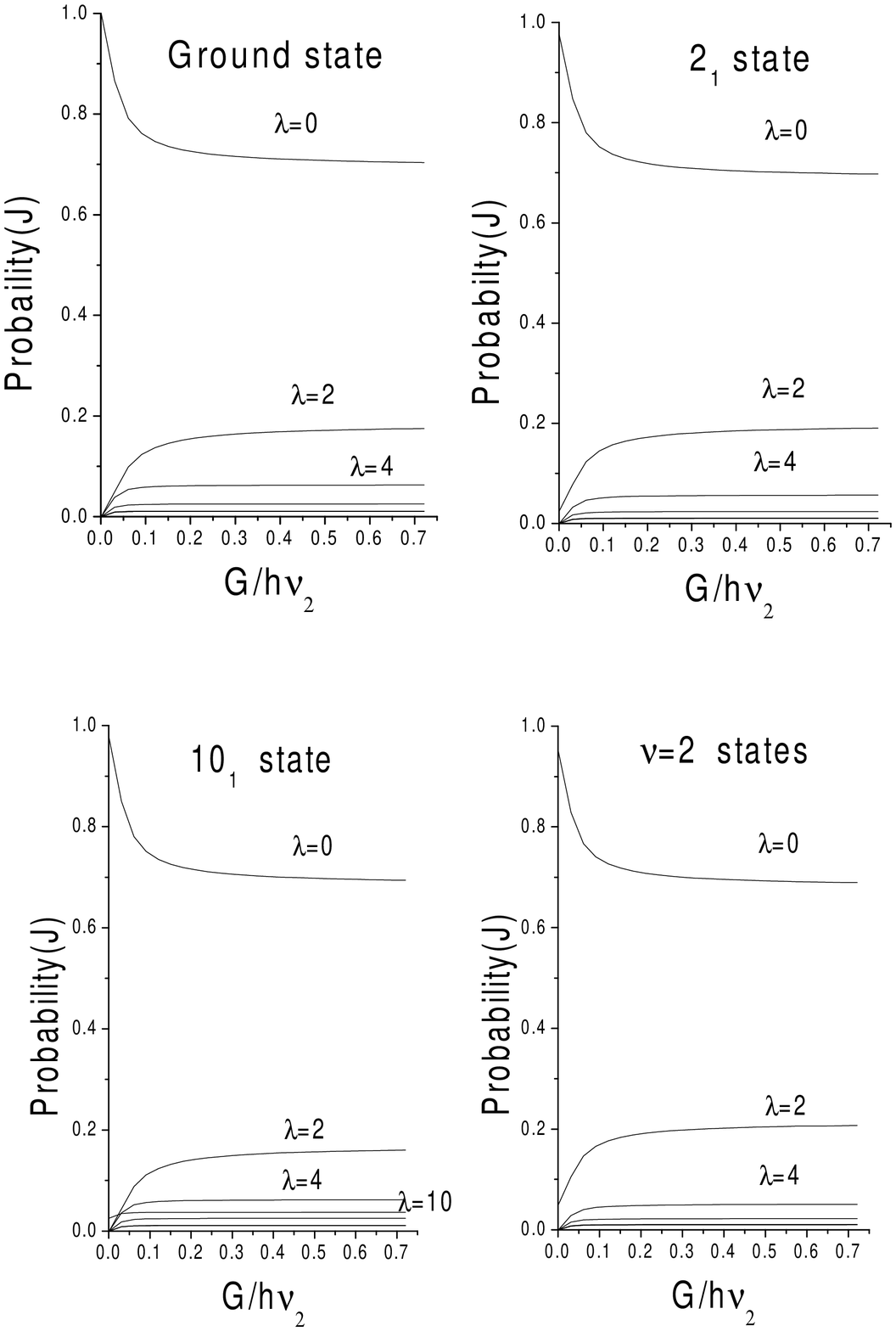}
\caption{Occupation probabilities of the different $\lambda$ bosons
as function of the scaled strength parameter $G/h\nu_2$ for a
calculation that includes all even values of $\lambda$ up to 12.
The lines corresponding to sizeable contributions are labelled by
$\lambda$.}
\label{fig3}
\end{figure}

In Table 2 we show the occupation probabilities for the different
boson degrees of freedom for $G/h\nu_2 =0.72$ and for the first state
corresponding to different seniorities. From this table follows
that the contributions of the different bosons (with seniority
one)are important in the description of the ground state band,
even if the occupation probabilities are small.\\

\begin{table*}[h*]\centering
\caption{Occupation probabilities for the bosons with angular
momentum J of the first states of a given seniority for the
calculation that includes all even J up to 12 with $G=0.72$. In
the last column are indicated the structure and angular momentums
of the corresponding states }
\begin{tabular}{cccccccc}
P(J=0) & P(J=2)& P(J=4)& P(J=6)& P(J=8) & P(J=10)& P(J=12) &
$States$\\
\tableline
 0.70384& 0.17515 &0.06277& 0.02505& 0.01092& 0.01104&
0.01124&
 J=0 ($\nu_i =0$)\\
0.69740& 0.19038& 0.05658& 0.02372& 0.01048& 0.01062& 0.01082& J=2
($\nu _{2}$=1)\\
0.69807& 0.14543 & 0.09903& 0.02464& 0.01079& 0.01092& 0.01112 &
J=4 ($\nu _{4}$=1)\\
0.69602& 0.15579& 0.06154 & 0.05365 &0.01085 & 0.01098 & 0.01118 &
J=6 ($\nu _{6}$=1)\\
0.69431 & 0.15963 & 0.06183& 0.02489 & 0.03715& 0.01099 & 0.01120&
J=8 ($\nu _{8}$=1)\\
0.69405 & 0.16008& 0.06186& 0.02489& 0.01087& 0.03704& 0.01120 &
J=10($\nu _{10}=1)$\\
0.69386 & 0.16040 & 0.06189 & 0.02490 & 0.01087&0.01100 &0.03710&
J=12($\nu _{12}=1)$\\
0.68952 & 0.20713 &0.05060& 0.02228& 0.00999& 0.01014& 0.01034&
J=2,4 ($\nu _{2}$=2)\\
0.68055& 0.22453&0.04514& 0.02082& 0.00964& 0.00964& 0.00984&
J=0,3,4,6 ($\nu _{2}=3)$\\
\end{tabular}
\label{table2}
\end{table*}

A question that one may ask is how will change the description
obtained as we diminished the number of boson degrees of freedoms.
For that purpose we also performed the calculation including only
bosons up to $\lambda=8$. We tried again to reproduce as well as
possible the rigid rotor by changing the bosons energies. The
values obtained for the energies (in the region of large G) were
E$_4=1.10$; E$_6=1.37$; E$_8=2.11$ which are similar to the values
used in the larger calculation. The asymptotic value for the $2^+$
energy was $0.101$.

\begin{figure}
\includegraphics[ height=6.5cm, width=5.2cm]{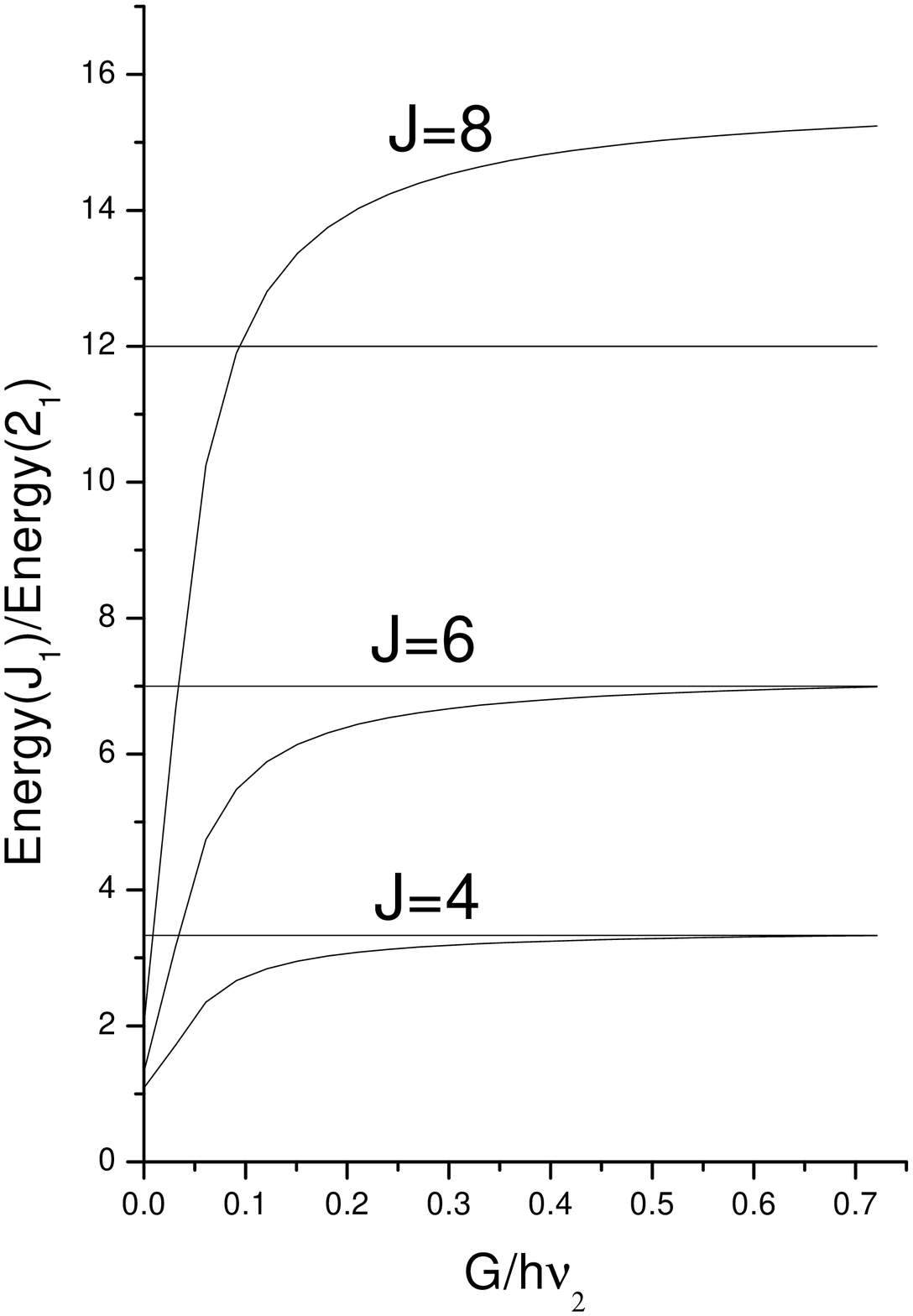}

\caption{Ratio of the energies of the $J_1$ states to
the energy of the $2_1$ state as function of the scaled strength
parameter $G/h\nu2$ for a calculation that includes all even
values of $\lambda$ up to 8. The rigid rotor values are shown to
guide the eye.}
 \label{fig4}
\end{figure}

In Fig. 4 we show results similar to Fig. 2 but using the
smaller space. It is clear that we are not able to reproduce
properly the $8^+$ state which gives an indication of the
importance of the $\lambda=10$ and $\lambda=12$ bosons in its
description. In Figure 5 we display the occupation probabilities
for this case and in Table 3 we reproduce the equivalent of Table 2.

\begin{table}[h]\centering
\caption{Occupation probabilities for the bosons with angular
momentum J of the first states of a given seniority for the
calculation that includes all even J up to 8 with $G=0.72$. In the
last column are shown the structure and angular momentums of
the corresponding states}
\begin{tabular}{cccccc}
P(J=0) & P(J=2)& P(J=4)& P(J=6)& P(J=8) &
$States$\\
\tableline
0.64192&0.22291&0.08728& 0.03680& 0.01108& J=0 ($\nu_i =0$)\\
0.63763& 0.23677& 0.07959& 0.03525& 0.01078& J=2 ($\nu _{2}$=1)\\
0.63747& 0.18653& 0.12862&0.03638&0.01100&J=4 ($\nu _{4}$=1)\\
0.63523&0.20049 & 0.08600& 0.06723&0.01105&J=6 ($\nu _{6}$=1)\\
0.63270& 0.20684& 0.08643& 0.03667& 0.03736 & J=8 ($\nu _{8}$=1)\\
0.63212&0.25220& 0.07181&0.03346&0.01041& J=2,4 ($\nu _{2}$=2)\\
0.62563&0.26826&0.06451&0.03158&0.01002&J=0,3,4,6 ($\nu
_{2}$=3)\\
\end{tabular}
\label{table3}
\end{table}

\begin{figure}
\includegraphics[ height=10.5cm, width=7.5cm]{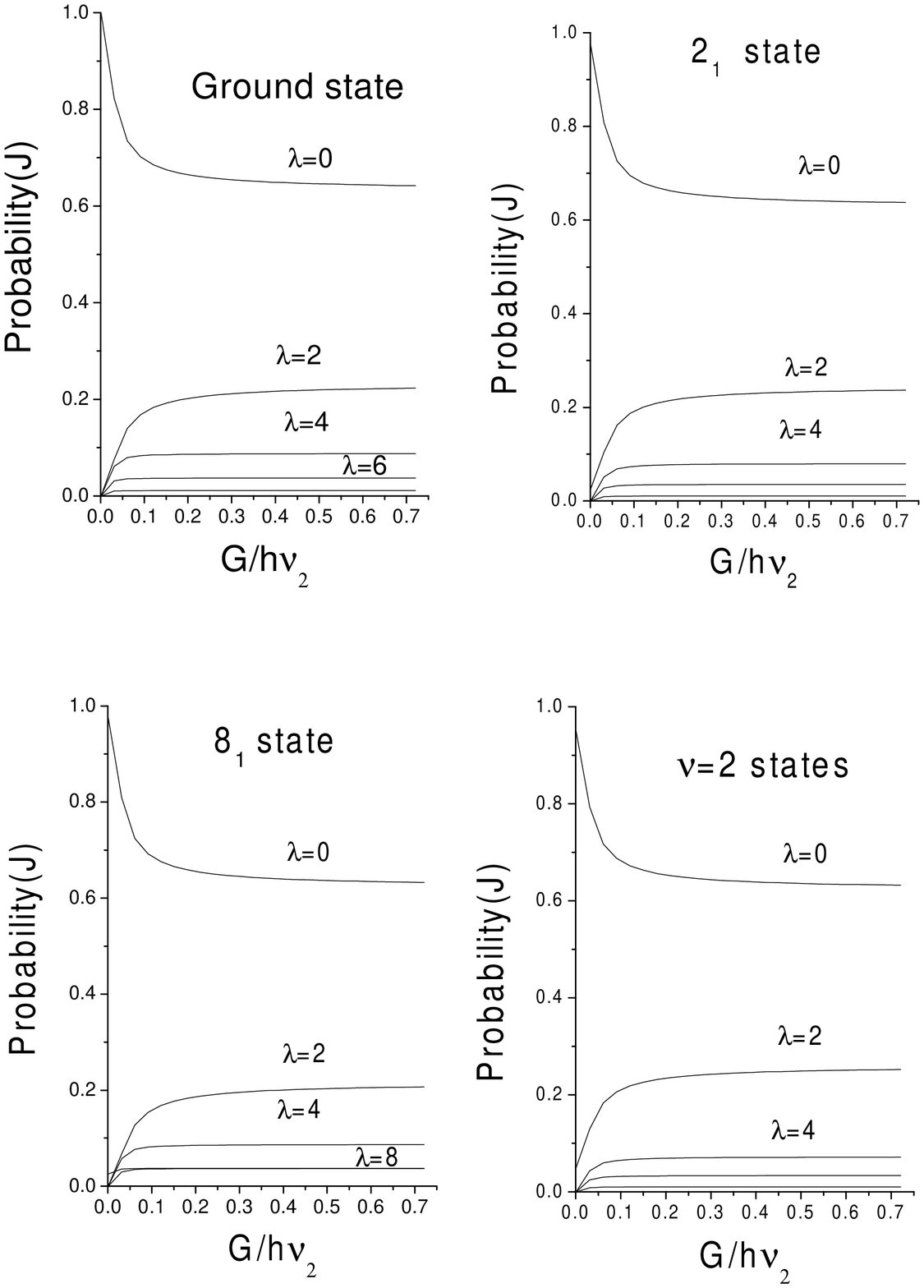}
\caption{Occupation probabilities of the
 different $\lambda$ bosons as function
of the scaled strength parameter $G/h\nu2$ for a calculation that
includes all even values of $\lambda$ up to 8. The lines are
labelled by $\lambda$.}
 \label{fig5}
\end{figure}
In Fig. 6 we show the energy ratios obtained for the ground
state band when we consider only S and D bosons. In this case
there are not free parameters (except for the pairing strength)
and it is quite clear that we are not able to reproduce a
rotational band.

\begin{figure}
\includegraphics[height=6.5cm, width=5.2cm]{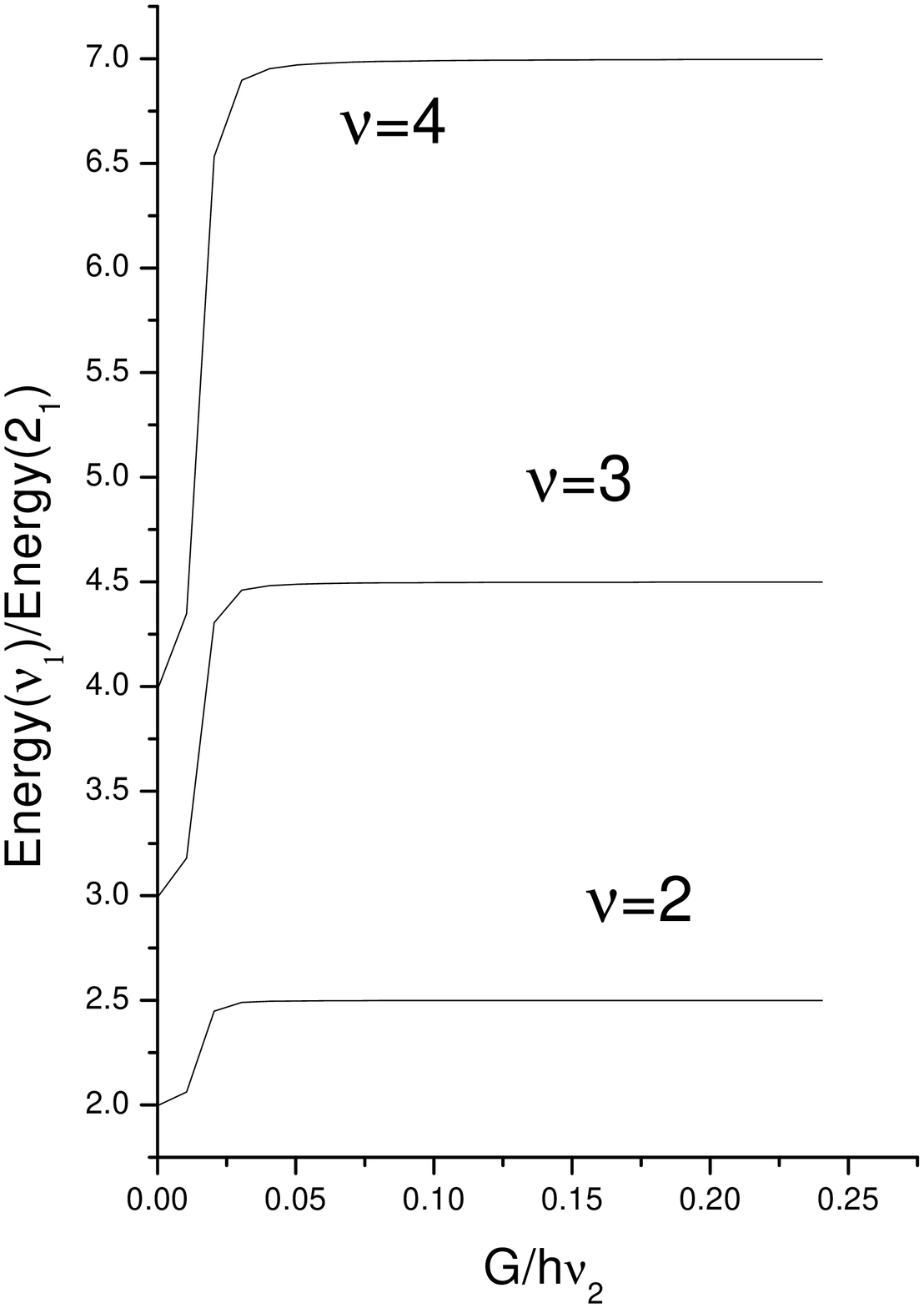}
\caption{Ratio of the energies of the states with
seniority $\nu$ to the energy of the $2_1$ state (corresponding to
seniority one) as function of the scaled strength parameter
$G/h\nu_2$ for a calculation that includes only S and D bosons.}
 \label{fig6}
\end{figure}

The main weakness for the relevance of the present calculation is
related to how well is taken into account the Pauli Principle by
our model hamiltonian. In a Nilsson like calculation one has two
main ingredients: on the one side the quadrupole-quadrupole
interaction deforms the systems, whose ground sate can be written
as a condensate of a boson that if formed by S, D, G, etc. bosons
with well-defined probabilities \cite{DDS1,DDS2} and then one must
consider the pairing interaction, that change by a small amount
these probabilities.  The contribution due to the pairing
interaction between fermions can be taken into account using Pauli
violating diagrams\cite{DDS1,DDS2}. The diagrams that appear in
this treatment are similar to the ones generated by our model
hamiltonian, but they do not have constant matrix elements. The
fact that one obtains a rotational spectra, when using a large
space, is an indication that the replacement of the correct matrix
element by a constant may not be a bad approximation.

We can now go on to the main conclusion of the present paper. The
S and D bosons used in the IBA model \cite{AI1,AI2} are different
from the ones that appears in a quadrupole-quadrupole+pairing
(Nilsson like) calculation. Using our model hamiltonian we have
shown that one cannot ignore the importance of the high angular
momentum bosons in the description of the ground sate
rotational band, even if the occupation probabilities of these
bosons are very small. Our result suggests that the S and D bosons
used in the IBA are not directly related to the ones obtained in a
Nilsson+BCS treatment. The S and D dominance obtained is a
manifestation of a general effect found in Ref. \cite{Rich}, but the
small amplitudes of the ``bosons'' with large angular momentum are
essential to obtain a good description of the ground state rotational band.

Valuable discussions with J. Dukelsky related to the work reported
herein are gratefully acknowledged.

This work was supported in part by the grant UBA-CYT x-204, x-053
and from CONICET PIP02618/00 as well as from Carrera del
Investigador Cient\'{\i}fico y T\'ecnico.

\section{Appendix 1 }
\centerline{{\large{\bf The single j shell}}}
\vspace{.5cm}

An attractive pairing-interaction between fermions in a single
j-shell has a well known exact solution that for the seniority zero
states is expressed in terms of the number of pairs of particles
(N) and the degeneracy ($\Omega=j+\frac{1}{2})$ as

\begin{equation}
E(N)=-|G|N\Omega \left(1-\frac{(N-1)}{\Omega}\right).
 \label{K}
\end{equation}

The second term can be considered to arise from the Pauli
principle between the pairs of fermions, each having an energy
$-|G|\Omega$.

In the case of S bosons the repulsive pairing hamiltonian is
written as ($G>0$)
\begin{equation}
H_{P}=W_s S^{\dagger}S+GS^{\dagger}S^{\dagger}SS,
 \label{H}
\end{equation}
and it is diagonal in the states of N bosons, yielding for the
energy
\begin{equation}
E(N)=W_s N+GN(N-1)=W_s N\left(1+\frac{G(N-1)}{W_s}\right).
 \label{F}
\end{equation}

To relate this result with the fermionic case one must use
$W_s=-|G|\Omega$ and then both energies coincide.

\clearpage


\begin{thebibliography}{}

\bibitem{DP}  J. Dukelsky and S. Pittel , Phys. Rev. Lett.{\bf  86}, 4791, 2001.

\bibitem{AI1}  A. Arima and F. Iachello, Ann. Phys., NY {\bf  99}, 253, 1976.

\bibitem{AI2}  A. Arima and F. Iachello, Ann. Phys., NY {\bf  111}, 20, 1978.

\bibitem{Ot1}  T. Otsuka, Nucl. Phys. {\bf  A368}, 244, 1981.

\bibitem{ZB}  M. R. Zirnbauer and D. M. Brink, Nucl. Phys. {\bf  A 384}, 1, 1982.

\bibitem{BM2}  C. Bloch and A. Messiah, Nucl. Phys. {\bf  9}, 95, 1962.

\bibitem{Gd1}  A. L. Goodman, Advances in Nuclear Structure, Vol. 11 
(ed. J.W.  Negele and E. Vogt, NY, Plenum)

\bibitem{Ot2}  T. Otsuka, A. Arima and N. Yoshinaga, Phys. Rev. Lett. 
{\bf  48}, 387, 1982.

\bibitem{DDS1}  J. Dukelsky, G. G. Dussel and H. M. Sofia, Nucl. Phys. 
{\bf  A373}, 267, 1982.

\bibitem{BM1}  A. Bohr and B. Mottelson, Phys. Scr. {\bf  22}, 468, 1980.

\bibitem{BBMV}  D. R. Bes, R. A. Broglia, E. Maglione and A. Vitturi, Phys.
Rev. Lett. {\bf  48}, 1001, 1982.

\bibitem{MVDB}  E. Maglione, A. Vitturi, C. A. Dasso R. A. Broglia, Nucl.
Phys. {\bf  A404}, 333,1983.

\bibitem{DDS2}  J. Dukelsky, G. G. Dussel and H. M. Sofia, J. Phys. G: Nucl.
Phys. {\bf  11},   L91, 1985.


\bibitem{DES1}  J. Dukelsky, C. Esebbag, and P. Schuck, Phys. Rev. Lett.
{\bf  87},  066403 (2001).

\bibitem{PDLD}  F. Pan, L-R. Dai, Y-A. Luo and J. P. Draayer, Phys. Rev. 
{\bf  C 68},  014308 (2003).

\bibitem{DPS}  J. Dukelsky, S. Pittel and G. Sierra, submitted to Reviews of
Modern Physics (2003).

\bibitem{Rich}  R. W. Richardson, Phys. Lett. {\bf  3},  277 (1963).

\bibitem{B5D}  D. R. Bes, Nucl. Phys. {\bf  10},  373 (1959).

\bibitem{Rich2} R. W. Richardson, J. Math. Phys. {\bf  9},  1327,
(1968).

\end{thebibliography}
\end{document}